\documentstyle[prl,aps,preprint]{revtex}

\let\ssection=\section
\renewcommand{\section}{\setcounter{equation}{0}\ssection}

\begin{document}
\draft
\title{An exact solution of the  metric-affine gauge theory with 
  dilation, shear, and spin charges} 

\author{Yu.N.\ Obukhov\footnote{Permanent address: Department of Theoretical
Physics, Moscow State University, 117234 Moscow, Russia}\and 
E.J.\ Vlachynsky\footnote{Permanent address: Department of Mathematics, 
University of Newcastle, Newcastle, NSW 2308, Australia}\and W.\ Esser\and 
R.\ Tresguerres\footnote{Permanent address: Consejo Superior de Investigaciones
Cientificas, Serrano 123, 28006 Madrid, Spain} and F.W.\ Hehl}
\address{Institute for Theoretical Physics,
University of Cologne\\D-50923 K{\"o}ln, Germany}

\maketitle
\bigskip
\bigskip

\begin{abstract}

  The spacetime of the metric-affine gauge theory of gravity (MAG)
  encompasses {\it nonmetricity} $Q_{\alpha\beta}$ and {\it torsion} 
  $T^\alpha$ as post-Riemannian structures. The sources of MAG are the 
  conserved currents of energy-momentum and dilation $\oplus$ shear 
  $\oplus$ spin. We present an exact static spherically symmetric vacuum
  solution of the theory describing the exterior of a lump of matter
  carrying mass and dilation $\oplus$ shear $\oplus$ spin charges.
\end{abstract}
\bigskip\bigskip
\bigskip

\section{Introduction}
\noindent
The four-dimensional affine group $A(4,R)$ is the semidirect product
of of the {\it translation} group $R^4$ and the {\it linear group}
$GL(4,R)$. If one gauges the affine group and additionally allows for
a metric $g$, then one ends up with a gravitational theory, the
metric-affine gauge theory of gravity (`metric-affine gravity' MAG),
see \cite{PR}, the spacetime of which encompasses two different
post-Riemannian structures: the nonmetricity one-form
$Q_{\alpha\beta}=Q_{i\alpha\beta} \,dx^i$ and the torsion two-form
$T^\alpha =\frac{1}{2}\,T_{ij}{}^\alpha dx^i\wedge dx^j$. Gauge models
in which $Q_{\alpha\beta}$ and $T^\alpha$ both provide propagating
modes, that is, if they are not tied to the material sources, have, in
the Yang-Mills fashion, gauge Lagrangians quadratic in curvature,
torsion, and nonmetricity:
\begin{equation} 
V_{\rm MAG}\sim\frac{1}{2\kappa}\,\left(R+\lambda +T^2+TQ+Q^2\right)
+R^2\,.\label{Vapprox}
\end{equation} 
Here we denote Einstein's gravitational
constant by $\kappa=\ell^2/(\hbar c)$ (the Planck length is $\ell$),
the cosmological constant by $\lambda$, and the curvature two-form by
$R_\alpha{}^\beta=\frac{1}{2}\,R_{ij\alpha}{}^\beta dx^i\wedge dx^j$.

One way to investigate the potentialities of such models is to 
look for {\it exact} solutions. Typically one starts with the 
Schwarzschild or the Kerr solution and tries to find a 
corresponding generalization appropriate for the spacetime 
geometry under consideration. As a first step, one 
requires the nonmetricity to vanish. Then the connection of 
spacetime is metric-compatible and the corresponding gauge model 
is described by the Poincar\'e gauge theory (PG) residing in a 
Riemann-Cartan spacetime. In the framework of the PG ($Q=0$), for a 
Lagrangian as in (\ref{Vapprox}), but without a linear piece $R$, 
McCrea et al., see \cite{McKerr}, found an electrically charged 
Kerr-NUT solution {\it with torsion}. Ordinary Newton-Einstein gravity 
is provided in such a model by the $T^2$-pieces, whereas the 
$R^2$-pieces supply `strong gravity' with a potential increasing 
as $r^2$, where $r$ is a radial coordinate. In a similar model, 
but with an external massless scalar field, Baekler et al.\cite{kink} 
found a torsion kink as an exact solution. Certainly, a lot still  
needs to be investigated in the framework of the PG, but the time 
seems ripe to turn one's attention to full-fledged MAG.

The search for exact solutions within MAG has been pioneered by
Tres\-guerres\cite{Tres14,Tres15} and by Tucker and Wang\cite{TW}.
With propagating nonmetricity $Q_{\alpha\beta}$ two types of charge are
expected to arise: {\it One dilation charge} related to the trace
$Q_\gamma{}^\gamma$ of the nonmetricity --- $Q:=Q_\gamma{}^\gamma/4$
is called the Weyl covector --- and {\it nine} types of {\it shear
  charge} related to the remaining traceless piece
${\nearrow\!\!\!\!\!\!\!Q}_{\alpha\beta}:=Q_{\alpha\beta}-Q\,g_{\alpha\beta}$ 
of the nonmetricity.  Under the
local Lorentz group, the nonmetricity can be decomposed into four
irreducible pieces $^{(I)}Q_{\alpha\beta}$, with $I=1,2,3,4$. It splits
according to $40=16\oplus 16\oplus 4\oplus 4$, see \cite{PR} p.122.
The Weyl covector is linked to $^{(4)}Q_{\alpha\beta}=Q\,g_{\alpha\beta}$. 
Therefore we should find $4+4+1$ shear charges and 1 dilation charge.

The simplest solution with nonmetricity should carry a dilation charge
(which couples to the Weyl covector $Q$) and should be free of shear
charges. Then the gauge Lagrangian $V_{\rm MAG}$ is not needed in its
full generality, rather we can restrict ourselves to
$\left(R+Q^2\right)/2\kappa +R^2$ or, more
exactly\cite{Palatini,Pono,TW}, with constants $\alpha$ and $\beta$, to
\begin{equation}
V_{\rm dil}=-{1\over2\kappa}\,\left(R^{\alpha\beta}\wedge
  \eta_{\alpha\beta}+ \beta\,Q\wedge{}^*\!\,Q\right) -{\alpha\over
  8}\,R_\alpha{}^\alpha\wedge{}^*\! R_\beta{}^\beta
\label{Vdil}\,.
\end{equation}
Here $\eta_{\alpha\beta}:={}^*\!\left(\vartheta_\alpha\wedge
  \vartheta_\beta\right)$, where $\vartheta^\gamma$ denotes the
coframe, and the star is the Hodge operator.  Observe that from the
nonmetricity {\it only the Weyl covector} $Q$ enters explicitly this
Lagrangian. The last term in (\ref{Vdil}) is proportional to the
square of Weyl's {\it segmental} curvature, $R_\alpha{}^\alpha=2d\,Q$.
This is a sheer post-Riemannian piece that would vanish identically in
any Riemannian spacetime.

{\it For $\beta=0$}, Tresguerres\cite{Tres14} found an exact dilation
solution which has a Reissner-Nordstr\"om metric, where the electric
charge is substituted by the dilation charge, a Weyl covector $\sim
1/r$, and a (constrained) torsion trace proportional to the Weyl
covector. Tucker and Wang\cite{TW} confirmed this result
independently. Later we will recover this dilation solution as a
specific subcase of our new solution.

The next step would then be to put on a proper shear charge, at least
one of the nine possible ones. From the point of view of the
irreducible decomposition of $Q_{\alpha\beta}$, the next simple thing,
beyond $^{(4)}Q_{\alpha\beta}=Q\,g_{\alpha\beta}$, is to pick the
vector piece $^{(3)}Q_{\alpha\beta}$, corresponding to {\it one} shear
charge.  We could now amend the Lagrangian (\ref{Vdil}) with terms of
the type $Q^{\alpha\beta}\wedge {}^*\!\, ^{(3)}Q_{\alpha\beta}$, but
it is more effective to turn immediately to a more general case.

\section{General quadratic MAG-Lagrangian}
\medskip

In a metric-affine space, the curvature has {\it eleven} irreducible
pieces, see \cite{PR}, Table~4.  If we recall that the nonmetricity
has {\it four} and the torsion {\it three} irreducible pieces
(loc.cit.), then a general quadratic Lagrangian in MAG reads
(signature $-+++$): 
\begin{eqnarray} 
\label{QMA} V_{\rm MAG}&=&
\frac{1}{2\kappa}\,\left[-a_0\,R^{\alpha\beta}\wedge\eta_{\alpha\beta}
-2\lambda\,\eta+ T^\alpha\wedge{}^*\!\left(\sum_{I=1}^{3}a_{I}\,^{(I)}
T_\alpha\right)\right.\nonumber\\
&+&\left.  2\left(\sum_{I=2}^{4}c_{I}\,^{(I)}Q_{\alpha\beta}\right)
\wedge\vartheta^\alpha\wedge{}^*\!\, T^\beta + Q_{\alpha\beta}
\wedge{}^*\!\left(\sum_{I=1}^{4}b_{I}\,^{(I)}Q^{\alpha\beta}\right)\right]
\nonumber\\&- &\frac{1}{2}\,R^{\alpha\beta} \wedge{}^*\!
\left(\sum_{I=1}^{6}w_{I}\,^{(I)}W_{\alpha\beta} +
  \sum_{I=1}^{5}{z}_{I}\,^{(I)}Z_{\alpha\beta}\right)\,.  
\end{eqnarray} 
In the above $\eta:={}^*\!\, 1$ is the volume four-form and the
constants $a_0,\cdots a_3$, $b_1,\cdots b_4$, $c_2, c_3,c_4$, $w_1,\cdots
w_6$, $z_1,\cdots z_5$ are dimensionless. In the curvature square term
we have introduced the irreducible pieces of the antisymmetric part
$W_{\alpha\beta}:= R_{[\alpha\beta]}$ and the symmetric part
$Z_{\alpha\beta}:= R_{(\alpha\beta)}$ of the curvature two-form.
Again, in $Z_{\alpha\beta}$, we meet a purely post-Riemannian part.
The segmental curvature is $^{(4)}Z_{\alpha\beta}:=
R_\gamma{}^\gamma\,g_{\alpha\beta}/4$.  In spite of this overabundance
of generality, it was possible \cite{Tres14,Tres15} to find shear type
solutions belonging to (\ref{QMA}), a search which we will continue
in this work.

To begin with, let us recall the three general field equations of MAG,
see \cite{PR} Eqs.(5.5.3)--(5.5.5). Because of its redundancy, we omit
the zeroth field equation with its gauge momentum $M^{\alpha\beta}$. The
first and the second field equations read 
\begin{eqnarray} 
DH_{\alpha}-
E_{\alpha}&=&\Sigma_{\alpha}\,,\label{first}\\ DH^{\alpha}{}_{\beta}-
E^{\alpha}{}_{\beta}&=&\Delta^{\alpha}{}_{\beta}\,,
\label{second}
\end{eqnarray} 
where $\Sigma_{\alpha}$ and $\Delta^{\alpha}{}_{\beta}$ are the
canonical energy-momentum and hypermomentum current three-forms
associated with matter. We will consider the {\it vacuum case} with
$\Sigma_{\alpha}=\Delta^{\alpha}{}_{\beta}=0$. The left hand sides of
(\ref{first})--(\ref{second}) involve the gravitational gauge field
momenta two-forms $H_{\alpha}$ and $H^{\alpha}{}_{\beta}$
(`excitations'). We find them, together with $M^{\alpha\beta}$, by
partial differentiation of the Lagrangian (\ref{QMA})
($T:=e_\alpha\rfloor T^\alpha\,,\> e_\alpha$ is the frame):
\begin{eqnarray}
M^{\alpha\beta}&:=&-2{\partial V_{\rm MAG}\over \partial Q_{\alpha\beta}}=
-{2\over\kappa}\Bigg[{}^*\! \left(\sum_{I=1}^{4}b_{I}{}^{(I)}
Q^{\alpha\beta}\right)\nonumber\\
&& + c_{2}\,\vartheta^{(\alpha}\wedge{}^*\! ^{(1)}T^{\beta)} +
c_{3}\,\vartheta^{(\alpha}\wedge{}^*\! ^{(2)}T^{\beta)} +
{1\over 4}(c_{3}-c_{4})\,g^{\alpha\beta}{}^*\!\,  T\Bigg]\,,\label{M1}\\
  H_{\alpha}&:=&-{\partial V_{\rm MAG}\over \partial T^{\alpha}} = -
  {1\over\kappa}\,
  {}^*\!\left[\left(\sum_{I=1}^{3}a_{I}{}^{(I)}T_{\alpha}\right) +
    \left(\sum_{I=2}^{4}c_{I}{}^{(I)}
      Q_{\alpha\beta}\wedge\vartheta^{\beta}\right)\right],\label{Ha1}\\
  H^{\alpha}{}_{\beta}&:=& - {\partial V_{\rm MAG}\over \partial
    R_{\alpha}{}^{\beta}}= {a_0\over 2\kappa}\,\eta^{\alpha}{}_{\beta} +
  {\cal W}^{\alpha}{}_{\beta} + {\cal Z}^{\alpha}{}_{\beta},\label{Hab1}
\end{eqnarray}
where we introduced the abbreviations
\begin{equation}
  {\cal W}_{\alpha\beta}:= {}^*\!
  \left(\sum_{I=1}^{6}w_{I}{}^{(I)}W_{\alpha\beta} \right),\quad\quad
  {\cal Z}_{\alpha\beta}:= {}^*\!
  \left(\sum_{I=1}^{5}z_{I}{}^{(I)}Z_{\alpha\beta} \right).
\end{equation}

Finally, the three-forms $E_{\alpha}$ and $E^{\alpha}{}_{\beta}$
describe the canonical energy-mo\-men\-tum and hypermomentum currents of
the gauge fields themselves. One can write them as follows \cite{PR}:
\begin{equation}
E_{\alpha}=e_{\alpha}\rfloor V_{\rm MAG} + (e_{\alpha}\rfloor T^{\beta})
\wedge H_{\beta} + (e_{\alpha}\rfloor R_{\beta}{}^{\gamma})\wedge
H^{\beta}{}_{\gamma} + {1\over 2}(e_{\alpha}\rfloor Q_{\beta\gamma})
M^{\beta\gamma},
\end{equation}
\begin{equation}
E^{\alpha}{}_{\beta}= - \vartheta^{\alpha}\wedge H_{\beta} - 
M^{\alpha}{}_{\beta}.
\end{equation}

\section{A spherically symmetric field configuration}

We take spherical polar coordinates $(t,r,\theta,\phi)$ and choose a
coframe of Schwarzschild type
\begin{equation}
  \vartheta ^{\hat{0}} =\, f\, d\,t \,,\quad\vartheta ^{\hat{1}} =\,
  {1\over f}\, d\, r\,, \quad\vartheta ^{\hat{2}} =\, r\,
  d\,\theta\,,\quad\vartheta ^{\hat{3}} =\, r\, \sin\theta \, d\,\phi
  \,,\label{frame2}
\end{equation}
with an unknown function $f(r)$. The coframe is assumed to be {\it
  orthonormal}, that is, with the local Minkowski metric
$o_{\alpha\beta}:=\hbox{diag}(-1,1,1,1) =o^{\alpha\beta}$, we have the 
special spherically symmetric metric 
\begin{equation} 
ds^2=o_{\alpha\beta}\,\vartheta^\alpha\otimes\vartheta^\beta 
= -f^2\,dt^2+\frac{dr^2}{f^2}
+r^2\left(d\theta^2+\sin^2\theta \,d\phi^2\right)\label{schwarz}\,.
\end{equation}

As for the torsion and nonmetricity configurations, we concentrate on
the simplest non-trivial case with shear. According to its irreducible
decomposition (see the Appendix B of \cite{PR}), the nonmetricity
contains two (co)vector pieces, namely $^{(4)}Q_{\alpha\beta}=
Q\,g_{\alpha\beta}$, the dilation piece, and
\begin{equation}
  ^{(3)}Q_{\alpha\beta}={4\over
    9}\left(\vartheta_{(\alpha}e_{\beta)}\rfloor \Lambda - {1\over
      4}g_{\alpha\beta}\Lambda\right)\,,\qquad \hbox{with}\qquad
  \Lambda:= \vartheta^{\alpha}e^{\beta}\rfloor\!
  {\nearrow\!\!\!\!\!\!\!Q}_{\alpha\beta}\label{3q}\,,
\end{equation}
a proper shear piece. Accordingly, our ansatz for the nonmetricity reads
\begin{equation}
  Q_{\alpha\beta}=\, ^{(3)}Q_{\alpha\beta} +\,
  ^{(4)}Q_{\alpha\beta}\,.\label{QQ}
\end{equation}
The torsion, in addition to its tensor piece,
encompasses a vector and an axial vector piece. Let us choose only
the vector piece as non-vanishing:
\begin{equation}
T^{\alpha}={}^{(2)}T^{\alpha}={1\over 3}\,\vartheta^{\alpha}\wedge T\,,
\qquad \hbox{with}\qquad T:=e_{\alpha}\rfloor T^{\alpha}\,.\label{TT}
\end{equation}

Thus we are left with the three non-trivial one-forms $Q$, $\Lambda$,
and $T$. In the spherically symmetric case, they must not distinguish
a direction in space. The following ansatz is compatible with that 
condition,
\begin{equation}
Q=u(r)\,\vartheta^{\hat{0}}\,,\quad\quad\Lambda=v(r)\,\vartheta^{\hat{0}}\,,
\quad\quad T=\tau(r)\,\vartheta^{\hat{0}}\,.\label{genEug}
\end{equation}

What are the consequences for the {\it curvature}? Let us examine the
zeroth Bianchi identity $DQ_{\alpha\beta}=2Z_{\alpha\beta}$. Its trace
is one irreducible piece (loc.cit.)  $2dQ=Z_\gamma{}^\gamma=
\,^{(4)}Z_\gamma{}^\gamma$, i.e., $Q$ serves as a {\it potential} for
$^{(4)} Z_\gamma{}^\gamma$, whereas its third part reads
$^{(3)}(DQ_{\alpha\beta})=2\,^{(3)}Z_{\alpha\beta}$, where
\begin{equation}
  ^{(3)}Z_{\alpha\beta}={2\over 3}\left(\vartheta_{(\alpha}\wedge
    e_{\beta)}\rfloor\Delta - {1\over
      2}g_{\alpha\beta}\Delta\right)\,,
  \quad\hbox{with}\quad\Delta:={1\over 2}\vartheta^{\alpha}\wedge
  e^{\beta}\rfloor\!{\nearrow\!\!\!\!\!\!\!Z}_{\alpha\beta}\,.\label{3z}
\end{equation}
The similarity in structure of (\ref{3q}) and (\ref{3z}) is apparent.
Indeed, {\it provided} the torsion carries only a vector piece, see
(\ref{TT}), we find 
\begin{equation}
\Delta={1\over 6}\,d\Lambda\,,\label{ddl}
\end{equation}
i.e.\ $^{(3)}Q_{\alpha\beta}$ acts as a potential for
$^{(3)}Z_{\alpha\beta}$.

\section{The new solution with four types of charge}

In the search for a solution we made heavy use of computer algebra. We
applied Reduce \cite{REDUCE} with its Excalc package \cite{EXCALC} for
treating exterior differential forms -- for sample programs see
\cite{Stauffer} -- and, furthermore, the Reduce-based GRG computer
algebra system \cite{GRG}. Nevertheless, the Lagrangian (\ref{QMA})
seemed to be too complicated. Having taken the lesson from the
dilation solution \cite{Tres14,TW}, we had to keep at least the square
of the segmental curvature in the Lagrangian. For simplicity, we
removed all other curvature square pieces by setting $w_{I}=0$ and
$z_{1}=z_{2}=z_{3}=z_{5}=0$. The Lagrangian then reduces to
\begin{eqnarray}
V&=& \frac{1}{2\kappa}\,\left[-a_0\,R^{\alpha\beta}
\wedge\eta_{\alpha\beta}-2\lambda\,\eta+ T^\alpha\wedge{}^*\!
\left(\sum_{I=1}^{3}a_{I}\,^{(I)}T_\alpha\right)\right.
\nonumber\\&+&\left.
2\left(\sum_{I=2}^{4}c_{I}\,^{(I)}Q_{\alpha\beta}\right)\wedge
\vartheta^\alpha\wedge{}^*\!\, T^\beta + Q_{\alpha\beta}\wedge{}^*\!
\left(\sum_{I=1}^{4}b_{I}\,^{(I)}Q^{\alpha\beta}\right)\right]
\nonumber\\&-& \frac{z_{4}}{2}\,R^{\alpha\beta} 
\wedge{}^*\!\,^{(4)}Z_{\alpha\beta}\,.\label{nondeg}
\end{eqnarray}

We substitute the local metric $o_{\alpha\beta}$, the coframe
(\ref{frame2}), the nonmetricity and torsion (\ref{genEug}) into the
field equations (\ref{first}), (\ref{second}) of the Lagrangian
(\ref{nondeg}). Then we find an exact solution with
\begin{equation}
f=\sqrt{1-{2\kappa M\over r} -{{\lambda}\,r^2\over {3a_0}} 
+z_4{{ \kappa(k_{0}{N})^2}\over{2a_0}\,r^2}}\label{f3}
\end{equation}and
\begin{equation}
u={k_{0}{N}\over f r}\,,\qquad v={k_{1}{N}\over f r}\,,\qquad \tau=
{k_{2}N\over f r}\,.\label{coul}
\end{equation}
Here $M$ and $N$ are arbitrary {\it integration constants}, and 
the coefficients $k_{0}, k_{1}, k_{2}$ are constructed in terms of the 
dimensionless coupling constants:
\begin{eqnarray}
k_0 &:=& \left({a_2\over 2}-a_0\right)(8b_3 + a_0) - 3(c_3 + a_0 )^2\,,
\label{k0}\\
k_1 &:=& -9\left[ a_0\left({a_2\over 2} - a_0\right) + 
(c_3 + a_0 )(c_4 + a_0 )\right]\,,
\label{k1}\\
k_2 &:=& {3\over 2} \left[ 3a_0 (c_3 + a_0 ) + (
8b_3 + a_0)(c_4 + a_0 )\right]\,.\label{k2}
\end{eqnarray}
A rather weak condition, which must be imposed on these coefficients, 
prescribes a value for the coupling constant $b_4$, namely
\begin{equation}
  b_4=\frac{a_0k+2c_4k_2}{8k_0}\,,\qquad\hbox{with}\qquad k:=
  3k_0-k_1+2k_2\,.\label{b4}
\end{equation}

If we collect our results, then the nonmetricity and the torsion read
as follows:
\begin{equation}
Q^{\alpha\beta}=\frac{1}{fr}\,\left[k_0N\,o^{\alpha\beta}+\frac{4}{9}\,
k_1N\,\left(\vartheta^{(\alpha}e^{\beta )}\rfloor-\frac{1}{4}\,
o^{\alpha\beta}\right)\right]\vartheta^{\hat{0}}\,,\label{nichtmetrizitaet}
\end{equation}
\begin{equation}
T^\alpha= \frac{k_2N}{3\,fr}\,\vartheta^\alpha\wedge\vartheta^{\hat{0}}
\,.\label{tosion}\end{equation}
Besides mass, this solution carries dilation, shear, and spin charges,
each of them of the (co)vectorial type. We have the following
assignments (the names in capitals are those from our computer programs):
\begin{eqnarray}\label{chargem}
M\>\; &\longrightarrow& \hbox{mass of Schwarzschild type}\,,\\ 
\label{chargek0}
k_0 N &\longrightarrow& \hbox{dilation (`Weyl') charge of type CONOM or
$^{(4)}Q^{\alpha\beta}$}\,,\\ \label{chargek1}
k_1 N &\longrightarrow& \hbox{shear charge of type VECNOM or 
$^{(3)}Q^{\alpha\beta}$}\,,\\ \label{chargek2}
k_2 N &\longrightarrow& \hbox{spin charge of type TRATOR or
$^{(2)}T^\alpha$}\,.
\end{eqnarray}      
For $N=0$ and $a_0=1$, we recover the Schwarzschild-deSitter solution
of general relativity.

\section{The curvature of the solution}

In order to verify this interpretation of the solution, we compute the
different curvature pieces.  For the post-Riemannian {\it symmetric}
part of the curvature we find:
\begin{equation}
{}^{(1)}Z^{\alpha\beta}={}^{(2)}Z^{\alpha\beta}=0\,,
\end{equation}
\begin{eqnarray}
{}^{(3)}Z^{\hat{0}\hat{0}}\,&=&\,-{}^{(3)}Z^{\hat{1}\hat{1}}\,=
\,{}^{(3)}Z^{\hat{2}\hat{2}}\,=\,
{}^{(3)}Z^{\hat{3}\hat{3}}\,=\,-{k_1 N\over 18r^2}
\,\vartheta^{\hat{0}}\wedge\vartheta^{\hat{1}},\nonumber\\
{}^{(3)}Z^{\hat{0}\hat{2}}\,&=&\,{k_1 N\over 18r^2}
\,\vartheta^{\hat{1}}\wedge\vartheta^{\hat{2}},\quad
{}^{(3)}Z^{\hat{0}\hat{3}}\,=\,{k_1 N\over 18r^2}
\,\vartheta^{\hat{1}}\wedge\vartheta^{\hat{3}},\nonumber \\ 
{}^{(3)}Z^{\hat{1}\hat{2}}\,&=&\,{k_1 N\over 18r^2}
\,\vartheta^{\hat{0}}\wedge\vartheta^{\hat{2}},\quad
{}^{(3)}Z^{\hat{1}\hat{3}}\,=\,{k_1 N\over 18r^2}
\,\vartheta^{\hat{0}}\wedge\vartheta^{\hat{3}},\label{3zstrich}
\end{eqnarray}
\begin{equation} 
^{(4)}Z^{\alpha\beta}=o^{\alpha\beta}\,{k_{0}N\over  2r^2}
\,\vartheta^{\hat{0}}\wedge\vartheta^{\hat{1}}.\label{z4strich}
\end{equation}
The components of $^{(3)}Z^{\alpha\beta}$ and $^{(4)}Z^{\alpha\beta}$,
the fields belonging to the potentials $\Lambda$ and $Q$, do {\it not}
depend on $M$, but rather only on the charges $k_1N$ and $k_0N$,
respectively.  In contrast,
\begin{eqnarray}
{}^{(5)}Z^{\hat{0}\hat{0}}\,&=&\,{}^{(5)}Z^{\hat{1}\hat{1}}\,=\,
                  -{k_1 N\over 9}\,\frac{h}{f^2r^2}
                  \,\vartheta^{\hat{0}}\wedge\vartheta^{\hat{1}},\quad
{}^{(5)}Z^{\hat{0}\hat{1}}\,=\,{kk_1 N^2\over 27f^2r^2}
                 \,\vartheta^{\hat{0}}\wedge\vartheta^{\hat{1}},\nonumber\\
{}^{(5)}Z^{\hat{0}\hat{2}}\,&=&\,\frac{k_1N}{9f^2r^2}\left({k N\over 3}
                      \, \vartheta^{\hat{0}}\wedge\vartheta^{\hat{2}}
                       +{h\over 2} 
                 \,\vartheta^{\hat{1}}\wedge\vartheta^{\hat{2}}\right),
                        \nonumber\\
{}^{(5)}Z^{\hat{0}\hat{3}}\,&=&\,\frac{k_1N}{9f^2r^2}\left({k N\over 3}
                      \, \vartheta^{\hat{0}}\wedge\vartheta^{\hat{3}}
                       +{h\over 2} 
                 \,\vartheta^{\hat{1}}\wedge\vartheta^{\hat{3}}\right),
                         \nonumber\\
{}^{(5)}Z^{\hat{1}\hat{2}}\,&=&\,-{k_1N\over 18}\,\frac{h}{f^2r^2}
                   \,\vartheta^{\hat{0}}\wedge\vartheta^{\hat{2}},\quad 
{}^{(5)}Z^{\hat{1}\hat{3}}\,=\,-{k_1N\over 18}\,\frac{h}{f^2r^2}
                   \, \vartheta^{\hat{0}}\wedge\vartheta^{\hat{3}},
\end{eqnarray}
where 
\begin{equation}
  h(r):=1 - {\lambda\, r^2\over a_0}-{\kappa\,q^2\over
    r^2}\qquad\hbox{and}\qquad q^2:={z_{4}\over 2a_{0}}\,
  (k_{0}N)^2\,,
\end{equation}
pick up $M$-dependent terms.

The {\it antisymmetric} or $SO(1,3)$ part of the curvature divides
into two classes: the class which may contain Riemannian pieces, namely 
the respective equivalents of the Weyl
tensor,
\begin{eqnarray} 
\!\!\!\!{}^{(1)}W^{\hat{0}\hat{1}}\!\!&=&\!-2\kappa\,{Mr - q^2\over r^4}
                  \,\vartheta^{\hat{0}}\wedge\vartheta^{\hat{1}},\quad
{}^{(1)}W^{\hat{2}\hat{3}}=-2\kappa\,{Mr - q^2\over r^4}
              \,\vartheta^{\hat{2}}\wedge\vartheta^{\hat{3}},\nonumber\\
{}^{(1)}W^{\hat{0}\hat{2}}\!\!&=&\!\quad\>\kappa\,{Mr - q^2\over r^4}
               \,\vartheta^{\hat{0}}\wedge\vartheta^{\hat{2}},\quad
{}^{(1)}W^{\hat{1}\hat{3}}=\quad\>\kappa\,{Mr - q^2\over r^4}
              \,\vartheta^{\hat{1}}\wedge\vartheta^{\hat{3}},\nonumber\\
{}^{(1)}W^{\hat{0}\hat{3}}\!\!&=&\!\quad\>\kappa\,{Mr - q^2\over r^4}
            \,\vartheta^{\hat{0}}\wedge\vartheta^{\hat{3}},\quad
{}^{(1)}W^{\hat{1}\hat{2}}=\quad\>\kappa\,{Mr - q^2\over r^4}
                  \,\vartheta^{\hat{1}}\wedge\vartheta^{\hat{2}},
\end{eqnarray}
the symmetric tracefree Ricci tensor, 
\begin{eqnarray}
{}^{(4)}W^{\hat{0}\hat{1}}\,&=&
\,\left({(\widetilde{k}N)^2\over f^2 r^2}+{\kappa q^2\over r^4}\right)
\vartheta^{\hat{0}}
                   \wedge\vartheta^{\hat{1}},\quad
  {}^{(4)}W^{\hat{2}\hat{3}}\,=\,-\left({(\widetilde{k}N)^2\over f^2 r^2}+
{\kappa q^2\over r^4}\right)
      \vartheta^{\hat{2}}\wedge\vartheta^{\hat{3}},\nonumber\\
  {}^{(4)}W^{\hat{0}\hat{2}}\,&=&
\quad{kN\over 12}\,\frac{h}{f^2 r^2}\,\vartheta^{\hat{1}}
\wedge\vartheta^{\hat{2}}+{(\widetilde{k}N)^2\over f^2 r^2}\,
\vartheta^{\hat{0}}\wedge
\vartheta^{\hat{2}},\nonumber\\ 
  {}^{(4)}W^{\hat{0}\hat{3}}\,&=&
\quad{kN\over 12}\,\frac{h}{f^2 r^2}\,\vartheta^{\hat{1}}\wedge
\vartheta^{\hat{3}}+{(\widetilde{k}N)^2\over f^2 r^2}\,
\vartheta^{\hat{0}}\wedge
\vartheta^{\hat{3}}\nonumber\\
{}^{(4)}W^{\hat{1}\hat{2}}\,&=&
\,-{kN\over 12}\,\frac{h}{f^2 r^2}\,\vartheta^{\hat{0}}\wedge
\vartheta^{\hat{2}}
 -{(\widetilde{k}N)^2\over f^2 r^2}\,\vartheta^{\hat{1}}\wedge
\vartheta^{\hat{2}},
\nonumber\\
  {}^{(4)}W^{\hat{1}\hat{3}}\,&=&
\,-{kN\over 12}\,\frac{h}{f^2 r^2}\,\vartheta^{\hat{0}}\wedge
\vartheta^{\hat{3}}
-{(\widetilde{k}N)^2\over f^2 r^2}\,\vartheta^{\hat{1}}\wedge
\vartheta^{\hat{3}},
\end{eqnarray}
where $\widetilde{k}^2:={1\over 648}\left({9k^2} + {4k_{1}^2}
\right)$,
and the curvature scalar,
\begin{equation}
{}^{(6)}W^{\alpha\beta}\,=\,\left(-{\lambda\over 3a_0} + 
{({4k_{1}^2} - {3k^2})\,N^2\over 216\,f^2 r^2} 
\right)\vartheta^{\alpha}\wedge\vartheta^{\beta}.
\end{equation}

The Weyl piece $^{(1)}W^{\alpha\beta}$ clearly displays the separation
into a Riemannian term containing the $M$ (corresponding to the
Schwarzschild solution) and the post-Riemannian $q^2$-term
proportional to the square of the dilation charge $k_0N$. The
symmetric tracefree Ricci piece $^{(4)}W^{\alpha\beta}$ represents the
`traceless energy-momentum' of the post-Riemannian charges. In the
scalar piece $^{(6)}W^{\alpha\beta}$ only the cosmological term
represents a Riemannian contribution.

The second class is the purely post-Riemannian class with 
\begin{equation}
{}^{(2)}W^{\alpha\beta}={}^{(3)}W^{\alpha\beta}=0\,,
\end{equation}
\begin{eqnarray}
  {}^{(5)}W^{\hat{0}\hat{2}}\,&=&\,{kN\over 12\,r^2}
     \,\vartheta^{\hat{1}}\wedge\vartheta^{\hat{2}},\quad
  {}^{(5)}W^{\hat{1}\hat{2}}\,=\,{kN\over 12\,r^2}
           \,\vartheta^{\hat{0}}\wedge\vartheta^{\hat{2}},\nonumber\\
  {}^{(5)}W^{\hat{0}\hat{3}}\,&=&\,{kN\over 12\,r^2}
         \,\vartheta^{\hat{1}}\wedge\vartheta^{\hat{3}},\quad
  {}^{(5)}W^{\hat{1}\hat{3}}\,=\,{kN\over 12\,r^2}
                     \,\vartheta^{\hat{0}}\wedge\vartheta^{\hat{3}}.
\end{eqnarray}
The components of $^{(5)}W^{\alpha\beta}$, corresponding to the antisymmetric
part of the Ricci tensor, do not depend on the mass $M$ at all.  For
the dilation solution \cite{Tres14,TW} we will find $k=0$, that is,
the antisymmetric Ricci piece will then vanish.

\section{Discussion}

With hindsight we observe that the constants $a_1\,,
a_3\,; b_1\,,b_2\,; c_2$ don't occur in our solution (\ref{f3}),
(\ref{coul}), nor in the constraint (\ref{b4}). In other words, {\it for
our solution} these constants are irrelevant and we can put them to zero,
\begin{equation} 
a_1 = a_3 = b_1 = b_2 = c_2 = 0.\label{constantszero}
\end{equation}
Then the Lagrangian (\ref{nondeg}) reduces to
\begin{eqnarray}
V&=& 
\frac{1}{2\kappa}\,\left[-a_0\,R^{\alpha\beta}\wedge
\eta_{\alpha\beta}-2\lambda\,\eta+
a_{2}\,T^\alpha\wedge{}^*\!\,^{(2)}T_\alpha\right.
\nonumber\\&+&\left.
2\left(c_{3}\,^{(3)}Q_{\alpha\beta}+c_{4}\,^{(4)}
Q_{\alpha\beta}\right)\wedge\vartheta^\alpha
\wedge{}^*\! T^\beta\right.\nonumber\\& +& 
\left.Q_{\alpha\beta}\wedge{}^*\!\left(b_{3}\,^{(3)}
Q^{\alpha\beta}+b_{4}\,^{(4)}Q^{\alpha\beta}\right)\right]
\nonumber\\&-& \frac{z_{4}}{2}\,R^{\alpha\beta} 
\wedge{}^*\!\,^{(4)}Z_{\alpha\beta}\,,\label{nondeg1}
\end{eqnarray}
where $b_4$ is determined by (\ref{b4}); in the term containing
${}^*\! T^\beta$ only ${}^*\!\,^{(2)} T^\beta$ survives. Consequently
{\it our solution} (\ref{frame2}), (\ref{schwarz}), (\ref{f3}),
(\ref{nichtmetrizitaet}), (\ref{tosion}), with (\ref{k0}), (\ref{k1}),
(\ref{k2}), (\ref{b4}), solves also the field equations belonging to
(\ref{nondeg1}). The structure emerging is transparent: In
(\ref{nondeg1}) nonmetricity and torsion enter explicitly only with
those pieces which are admitted according to (\ref{chargek0}),
(\ref{chargek1}), (\ref{chargek2})!

Since the dilation solution of Tresguerres \cite{Tres14} carries no
shear charge $k_1N$ and its antisymmetric Ricci is identical zero, we
can recover it by putting $k_1=0$ and $k=0$. This yields two
underdetermined equations. A particular solution can be obtained by
means of the ansatz $b_3=c_3=0$. Then we find $c_4 =-{1\over 2}\,a_2$,
$\;k_0 = -{2\over 3}\,k_2 = a_{0}\,({1\over 2}\,a_2 - 4a_{0})$, and
$b_4=\frac{3}{16}\,a_2$. This corresponds, up to an exact form, to the
Lagrangian of \cite{Tres14} Eq.(3.23).

Let us return to the dilation Lagrangian (\ref{Vdil}). We would like
to generalize it in a minimal way such that it allows for our
solution. Taking in (\ref{nondeg1}) $b_3=c_3=0$, as above, and putting
$c_4=0$ and $a_0=1,\,\lambda=0$, we find
\begin{equation}
  V_{\rm dil-sh}=-{1\over2\kappa}\,\left(R^{\alpha\beta}\wedge
    \eta_{\alpha\beta}+
    \beta\,Q\wedge{}^*\!\,Q+\gamma\,T\wedge{}^*\!\,T\right)
  -{\alpha\over 8}\,R_\alpha{}^\alpha\wedge{}^*\! R_\beta{}^\beta
\label{Vdil-sh}\,,
\end{equation}
with $\alpha:=z_4,\,\beta:=-4b_4,\,\gamma:=-\frac{1}{3}\,a_2$. In
terms of these parameters, the $b_4$-constraint (\ref{b4}) reads:
\begin{equation}
\gamma=-\frac{8}{3}\;\frac{\beta}{\beta+6}\,.\label{constraint}
\end{equation}
The post-Riemannian fields (\ref{coul}), with a suitably redefined
integration constant $\widetilde{N}:=-24{N}/({\beta-6})$,
turn out to be
\begin{equation}
  u=\,{\widetilde{N}\over f
    r}\,,\qquad v=\frac{3\beta}{2}\;{\widetilde{N}\over f
    r}\,,\qquad \tau=
  -\left(\frac{3}{2}+\frac{\beta}{4}\,\right)\,{\widetilde{N}\over f
    r}\,.\label{coul1}
\end{equation}
Thus our solution fulfills the field equations of (\ref{Vdil-sh}),
provided (\ref{constraint}) is valid. The prescription $\beta=0$, via
(\ref{constraint}), implies $\gamma=0$, and the post-Riemannian fields
(\ref{coul1}) collapse to the dilation subcase of
Tresguerres\cite{Tres14} and Tucker and Wang\cite{TW}.

If, instead of a Reissner-Nordstr\"om type ansatz, we started with one
using the Kerr-NUT-Newman solution, then we would be able to find
axially symmetric solutions carrying nonmetricity and torsion. However,
we will leave that to a subsequent publication.

\noindent{\bf Acknowledgments}
This research was supported by the Deutsche Forschungsgemeinschaft (Bonn)
under project He-528/17-1 and by the Graduate College ``Scientific 
Computing'' (Cologne-St.Augustin).

\bigskip

\end{document}